\documentclass[useAMS,usenatbib]{mn2e}
\usepackage{rotating}
\usepackage{longtable}
\usepackage{graphics}
\usepackage{times}
\usepackage[usenames,dvipsnames]{xcolor}

\def\teff{$T_{\rm eff}$}
\def\ms{m\,s$^{-1}$}
\def\kms{km\,s$^{-1}$}
\def\i{\,{\sc i}}
\def\ii{\,{\sc ii}}
\def\iii{\,{\sc iii}}
\def\ha{H$\alpha$}

\def\vs{$v_{\rm e}\sin i$}
\def\bb{$\langle B \rangle$}
\def\bz{$\langle B_{\rm z} \rangle$}
\def\synthmag{{\sc SynthMag}}
\def\lgg{$\log g$}
\def\ha{H$\alpha$}

\def\aj{AJ}%
\def\apj{ApJ}%
\def\apjl{ApJ}%
\def\aap{A\&A}%
\def\aaps{A\&AS}%
\def\mnras{MNRAS}%
\newcommand{\bs}{$\langle B \rangle$}

\newcommand{\fifps}[2]{\centering\resizebox{#1}{!}{\includegraphics{#2}}}

\begin{document}

\title[New roAp stars]
{Discovery of new roAp pulsators in the UVES survey of cool magnetic Ap stars%
\thanks{Based on observations collected at the European Southern Observatory, Paranal, Chile (program 085.D-0124).}
}

\author[O.~Kochukhov et al.]%
{O.~Kochukhov$^1$,
D.~Alentiev$^{2,3}$,
T.~Ryabchikova$^4$,
S.~Boyko$^5$,
M.~Cunha$^3$, 
V.~Tsymbal$^2$, \and 
W.~Weiss$^6$ \\
$^1$ Department of Physics and Astronomy, Uppsala University Box 516, 751 20 Uppsala, Sweden\\
$^2$ Department of Physics, Tavrian National University, Vernadskiy's Avenue 4, 95007 Simferopol, Ukraine\\
$^3$ Centro de Astrofisica da Universidade do Porto, Rua das Estrelas, 4150-762 Porto, Portugal\\
$^4$ Institute of Astronomy, Russian Academy of Sciences, Pyatnitskaya 48, 119017 Moscow, Russia \\
$^5$ Department of Physics, M.V.Lomonosov Moscow State University, GSP-1, 1-2 Leninskye Gory, 119991 Moscow, Russia \\ 
$^6$ Department of Astronomy, University of Vienna, T\"urkenschanzstrasse 17, 1180 Wien, Austria}

\date{Accepted 2013 February 26.  Received 2013 February 7; in original form 2012 December 19}

\pagerange{\pageref{firstpage}--\pageref{lastpage}}
\pubyear{2012}

\maketitle

\label{firstpage}

\begin{abstract}
We have carried out a survey of short-period pulsations among a sample of carefully chosen cool Ap stars using time-resolved observations with the UVES spectrometer at the ESO 8-m VLT telescope. Here we report the discovery of pulsations with amplitudes 50--100~\ms\ and periods 7--12~min in HD\,132205, HD\,148593 and HD\,151860. These objects are therefore established as new rapidly oscillating Ap (roAp) stars. In addition, we independently confirm the presence of pulsations in HD\,69013, HD\,96237 and HD\,143487 and detect, for the first time, radial velocity oscillations in two previously known photometric roAp stars HD\,119027 and HD\,185256. At the same time, no pulsation variability is found for HD\,5823, HD\,178892 and HD\,185204. All of the newly discovered roAp stars were previously classified as non-pulsating based on the low-precision ground-based photometric surveys. This shows that such observations cannot be used to reliably distinguish between pulsating and non-pulsating stars and that all cool Ap stars may harbor \textit{p}-mode pulsations of different amplitudes.
\end{abstract}


\section{Introduction}
\label{intro}

Rapidly oscillating Ap (roAp) stars are unique astrophysical laboratories allowing the study of the effects that strong organised magnetic fields have on stellar rotation, convection, pulsations and chemical element transport in the stellar interiors and atmospheres. These stars belong to the group of chemically peculiar, magnetic late-A and early-F objects, commonly known as SrCrEu Ap stars. The roAp stars exhibit high-overtone, non-radial \textit{p-}mode pulsations with periods around 10~min and low amplitudes both in photometry and spectroscopy \citep{kurtz:2000,kochukhov:2008c}. The presence of multi-periodic pulsations in many roAp stars makes them interesting targets for a classical asteroseismic analysis concerned with determining global stellar properties \citep[e.g.,][]{saio:2010}. In addition, spectroscopic observations of the variability in rare-earth spectral lines formed in the outer atmospheric layers of these stars offer unique possibilities for tomographic mapping of the vertical structure of pulsation modes \citep{ryabchikova:2007b} and for investigating intricate details of the physics of propagating magneto-acoustic waves \citep{khomenko:2009}. 

Currently, we have only a limited understanding of the physical processes responsible for the excitation of high-overtone \textit{p-}mode oscillations in magnetic Ap stars. The most plausible theory \citep{balmforth:2001} predicts that suppression of convection in the outer stellar layers allows excitation of the roAp pulsations due to the $\kappa$-mechanism operating in the hydrogen ionization zone. Since the driving of the oscillations results from a subtle energy balance that depends directly on the interaction between the magnetic field, convection, pulsations, and atomic diffusion, pulsational analysis provides a unique tool for studying these physical processes and their coupling.  

Modern theoretical pulsation models are fairly successful in matching the observed pulsation frequencies \citep{kurtz:2002,mkrtichian:2008} and even in assessing a complex geometry of pulsation modes distorted by rotation and magnetic field \citep{saio:2004,kochukhov:2004f,bigot:2011}, but are less predictive when it comes to explaining distribution of the roAp stars in the H-R diagram. Compared to observations, excitation models  predict pulsations in systematically hotter and more luminous Ap stars \citep{cunha:2002,theado:2009}. In fact, a significant fraction of more than 40 currently known roAp stars are located beyond the red boundary of the theoretical instability strip.

A related observational difficulty is the co-existence of pulsating and apparently constant Ap stars in the same region of the H-R diagram. The separation between the roAp and non-pulsating Ap (noAp) stars is largely reliant on the historic ground-based surveys \citep[e.g.,][]{nelson:1993,martinez:1994}, which might not be sensitive enough to reveal low-amplitude photometric variability. In fact, observations with the Kepler satellite have shown that in some roAp stars the amplitude of the oscillations does not exceed a few tens of $\mu$mag \citep{balona:2011}. Clearly such roAp stars would have been identified as constant from ground-based observations.

Recent spectroscopic detections of pulsations in some of the prototypical ``photometric noAp'' stars \citep{hatzes:2004,elkin:2005,kochukhov:2009} demonstrated the clear advantages of the spectroscopic observations over ground-based photometry in discovering and characterizing pulsations in cool Ap stars. In particular, high-resolution spectroscopy allows one to isolate rare-earth lines, which often show 10--100 times higher pulsational amplitudes than the lines of light and iron-peak elements. 

Aiming to establish an unbiased incidence of rapid oscillations among cool Ap stars, we have carried out a survey of pulsations in a small sample of roAp-candidates using the most powerful instrumentation currently available for high-resolution stellar spectroscopy. Our observations, performed with the UVES spectrometer at the ESO VLT telescope, turned out to be remarkably successful as we were able to demonstrate the presence of pulsations in 9 out of 12 observed stars. The discovery of the longest-period roAp star, HD\,177765, was reported in a separate paper \citep{alentiev:2012}. Here we present further discoveries of new roAp stars, confirmations of previously known ones and report the sensitive upper limits on the radial velocity pulsations for a few objects in which we could not detect variability. 

This paper is structured as follows. Target selection is discussed in Sect.~\ref{targets}. Observations and data reduction are outlined in Sect.~\ref{observ}. Details of the methods employed for radial velocity analysis and atmospheric parameter determination of the target stars are given in Sect.~\ref{rvmethod} and \ref{params} respectively. Results for individual stars are presented in Sect.~\ref{results}. The outcome of our survey is summarised and discussed in the context of other recent studies of the roAp stars in Sect.~\ref{conclusion}.

\section{methods}
\label{methods}

\subsection{Target selection}
\label{targets}

The primary list of roAp candidates was drawn from the catalogue by \citet{renson:2009}, taking into account a number of recent publications on individual Ap stars. We have selected Ap stars falling below the $T_{\rm eff}$\,=\,8000~K threshold beyond which few roAp stars are observed. We have also given preference to observe objects with reasonably sharp spectral lines, for which best accuracy in radial velocity measurements can be expected. ESO archival spectra, such as those described by \citet{freyhammer:2008}, were extensively used to confirm the cool Ap-star nature of the targets. We found that many of the late-A objects classified as, e.g. ``Ap Sr'' by \citet{renson:2009} and included in previous roAp photometric surveys, are either Am stars or very rapid rotators for which spectral classification is ambiguous. Finally, we compiled a list of 14 stars, 12 of which were eventually observed during ESO Period 85. This sample included two known photometric roAp stars, HD\,119027 and HD\,185256, for which no time-resolved spectroscopic studies have been previously carried out.

\subsection{Observations and data reduction}
\label{observ}

The candidate and known roAp stars were observed in the period from April to July 2010 using the Ultraviolet and Visual Echelle Spectrograph (UVES) installed at one of the ESO 8-m VLT telescopes. The spectrograph was configured to use the 600~nm red setting with an image slicer. This setup provided resolution $R\approx110\,000$ and the wavelength coverage from 4980~\AA\ to 7010~\AA\ with a 100~\AA\ gap in the region centered at 5990~\AA. 

Each stellar observation consisted of 50--67 exposures. Individual exposure time varied between 60 and 90~s, depending on the stellar brightness. To optimise these time-series observations we employed the ultra-fast (4-port, 625~kpix\,s$^{-1}$) readout mode of the UVES CCDs, which allowed us to reduce the overhead between consecutive exposures to 21~s.

In total, we obtained 716 spectra with a signal-to-noise ratio of 40--160 over 10 observing nights. The known roAp star HD\,119027 was observed on two occasions, with 50 exposures each. For another target, HD\,185204, an incomplete time-series of 38 exposures was obtained in addition to the standard 50-exposure sequence. We analysed these multiple datasets available for HD\,119027 and HD\,185204 individually.
 
Reduction of the echelle spectra was carried out with an improved version of our UVES pipeline \citep{alentiev:2012}. This code preforms common reduction steps, such as bias and scattered light subtraction, order position determination, extraction of one-dimensional spectra, wavelength calibration and continuum normalization. A barycentric radial velocity corrections was taken into account in the final reduction step.

Detailed information on the adopted exposure times, the number of spectra obtained, the Julian date of the start and end of observations, and typical signal-to-noise ratios is given in Table\,\ref{obs-table}.

\begin{figure*}
\centering
\fifps{15cm}{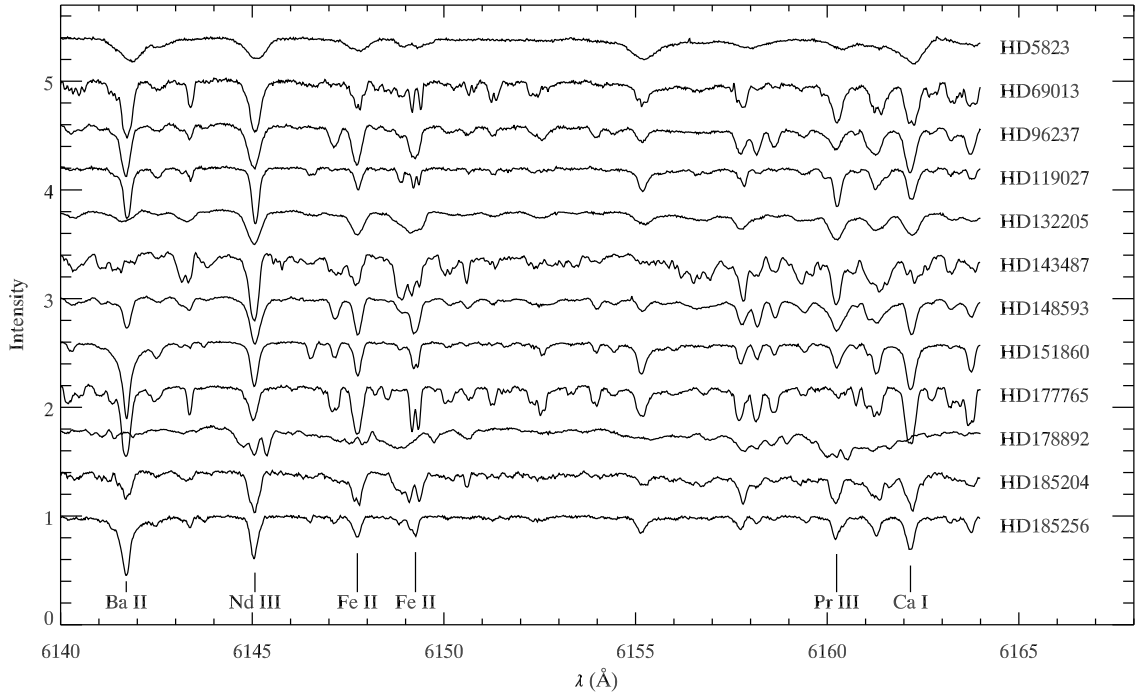}
\caption{Average spectra of program stars in the 6140--6164~\AA\ region. For completeness we also show the mean spectrum of HD\,177765 studied by \citet{alentiev:2012}. Identification is provided for most important spectral features.}
\label{allspectra}
\end{figure*}

\begin{table*}%[!t]
\caption{The journal of UVES observations of roAp stars and roAp candidates. The columns give the star name, the magnitude in Johnson $V$ band, the number of spectra, individual exposure times, the start and end heliocentric Julian dates of time-series observations, and the typical signal-to noise ratio. Superscripts denote data sets obtained on different observing nights for HD\,119027 and HD\,185204.
\label{obs-table}
}
\centering
\begin{tabular}{lcccccc}
\hline
\hline
Star & $V$& $N$ & $T_{\rm exp}$ (s) & HJD$_{\rm start}-2455000$ & HJD$_{\rm end}-2455000$ & $S/N$\\
\hline
HD 5823      & 9.98& 50 & 80 & 372.88456 & 372.94191 & 60--100\\
HD 69013     & 9.56&62 & 60 & 300.46474 & 300.52203 & 40--70\\
HD 96237     & 9.45&50 & 60 & 300.53081 & 300.57686 & 70--115\\
HD 119027$^1$& 9.92&50 & 90 & 289.82059 & 289.88364 & 55--90\\
HD 119027$^2$& 9.92&50 & 90 & 300.58321 & 300.64623 & 45--75\\
HD 132205    & 8.72&50 & 60 & 288.85571 & 288.90175 & 100--160\\
HD 143487    & 9.43&62 & 60 & 374.71393 & 374.77122 & 70--110\\
HD 148593    & 9.15&50 & 60 & 389.67704 & 389.72305 & 80--130\\
HD 151860    & 9.01&50 & 60 & 326.83944 & 326.88543 & 85--130\\
HD 178892    & 8.94&67 & 60 & 351.83831 & 351.90025 & 80--130\\
HD 185204$^1$& 9.53&38 & 60 & 349.64696 & 349.68171 & 45--70 \\
HD 185204$^2$& 9.53&50 & 60 & 382.88979 & 382.93578 & 45--75\\
HD 185256    & 9.96&50 & 80 & 324.79581 & 324.85315 & 45--75\\
\hline
\end{tabular}
\end{table*}

\subsection{Radial velocity analysis}
\label{rvmethod}

Reduced, one-dimensional spectra were used to determine radial velocity (RV) variation in individual lines and groups of lines. The atomic data necessary for the identification of spectral lines were obtained from the VALD database \citep{kupka:1999} and adopted from our previous studies of the roAp stars \citep[e.g.,][]{ryabchikova:2007a}. We made an effort to select only spectral lines not significantly affected by blends.

The RV analysis started with determinations of the line centers using the center-of-gravity technique described by \citet{kochukhov:2001b}. With our data quality, individual spectral lines seldomly provided accurate enough RVs for an unambiguous detection of pulsations. Therefore, we constructed average RV curves for all suitable lines of a given ion by removing occasional linear trends from the RV data of individual lines and averaging the resulting curves. Then, discrete Fourier transform was used to obtain an amplitude spectrum and derive an initial guess for the pulsation period from the position of the highest peak. The False Alarm Probability (FAP) of this variable signal was estimated as $FAP=1- [1-\exp{(-z)}]^N$, where $z$ is the height of the peak in the variance-normalised Lomb-Scargle periodogram \citep{scargle:1982} and the number of independent frequencies $N$ was computed according to the prescription given by \citet{horne:1986}. The mean pulsation period was determined by averaging period estimates for all ions with definite detection of variability (FAP\,$\le$\,$10^{-3}$). In the final step, RV curves of individual ions were fitted with a linear least-squares algorithm to obtain an estimate of the pulsation phase and amplitude. 

Fitting a cosine curve to the RV data allows us to obtain a proxy of the relative formation height of variable spectral lines. As shown by \citet{ryabchikova:2007b}, many roAp stars exhibit a regular progression in the phase of pulsation maximum from one ion to the next. This picture reflects the outward propagation of pulsation waves through the chemically stratified stellar atmosphere. As a first approximation, one can assume that a smaller phase in the cosine function corresponds to a later pulsation maximum and hence to the line formation in the higher atmospheric layers and vice versa.

In addition to looking at the lines of rare-earth ions typically showing the largest amplitudes for the roAp stars, we also carried out a frequency analysis of mean RV curves of a non-variable ion, typically Fe\i, to assess intrinsic stability of the spectrograph in the course of our observations and detect possible spurious variability due to instrumental artifacts.

\begin{table}%[!t]
\caption{Basic parameters of the program stars. The columns give the star name, spectral classification according to \citet{renson:2009} and estimates of \teff, \lgg, \vs\ and mean magnetic field modulus \bb. Asterisk marks magnetic field modulus determined with synthetic spectrum modeling. 
\label{params-table}
}
\centering
\begin{tabular}{llcccc}
\hline
\hline
Star & \multicolumn{1}{c}{Spectral} & \teff\ & \lgg\ & \multicolumn{1}{c}{\vs}  & \bb\ \\
 & \multicolumn{1}{c}{type} & (K) & & (\kms) & (kG)\\
\hline
HD 5823   & F2 SrEuCr & 7300   & 4.3 & 13.5      & 8.5$^\ast$ \\ % TR syn
HD 69013  & A2 SrEu   & 7600   & 4.5 & 4.0       & 4.8\phantom{$^\ast$} \\ % OK 4842
HD 96237  & A4 SrEuCr & 7800   & 4.3 & 6.0       & 2.9$^\ast$ \\ % TR syn
HD 119027 & A3 SrEu   & 7050   & 4.4 & 4.0       & 3.1\phantom{$^\ast$} \\ %OK 3148 (A), 3141 (B)
HD 132205 & A2 EuSrCr & 7800   & 4.4 & 9.5       & 5.2$^\ast$ \\ %syn of the magnetically splitted Fe I 6336 line 
HD 143487 & A3 SrEuCr & 7000   & 5.0 & 1.5       & 4.7\phantom{$^\ast$} \\ %4.75$\pm$0.06(TR), 4027(AD), (OK)
HD 148593 & A2 Sr     & 7850   & 4.4 & 5.0       & 3.0$^\ast$ \\ %syn? syn width of Fe II 6149, Fe I 6336
HD 151860 & A2 SrEu   & 7050   & 4.5 & 4.5       & 2.5\phantom{$^\ast$} \\ % AD, OK 2529
HD 178892 & Ap SrCrEu & 7700   & 4.0 & 10.0      & 18.5$^\ast$ \\ %TR by Nd 3 6690, Pr 3 6706,
HD 185204 & A2 SrEuCr & 7750   & 4.4 & 4.5       & 5.4\phantom{$^\ast$} \\ %AD OK 5407
HD 185256 & F0 SrEu   & 7150   & 4.3 & 5.5       & $\le$1.4$^\ast$ \\ % TR syn
\hline
\end{tabular}
\end{table}

\subsection{Stellar parameters}
\label{params}

Effective temperatures of the target stars were estimated using the Str\"{o}mgren photometric data from \citet{martinez:1993} and applying calibrations by \citet{moon:1985} and \citet{napiwotzki:1993} as implemented in the {\sc templogg} code \citep{kaiser:2006}. For HD\,178892, which lacks the Str\"omgren photometry, we adopted atmospheric parameters from \citet{ryabchikova:2006a}.

Measurements of the mean magnetic field modulus were accomplished by Gaussian fitting the resolved components of the Zeeman split Fe\ii\ 6149.26~\AA\ line. In the cases when this line did not show Zeeman splitting or was distorted by blends, we applied Gaussian fitting to a few other spectral features with simple Zeeman splitting patterns or used the \synthmag\ polarised spectrum synthesis code \citep{kochukhov:2007d} to model magnetic broadening of magnetically sensitive spectral lines. The latter procedure was also necessary when the lines were dominated by rotational broadening. The accuracy of the magnetic field estimated by fitting well-resolved Zeeman split lines is about 0.1~kG. On the other hand, the spectrum synthesis determination of the magnetic field modulus is accurate to within $\sim$\,0.5~kG. \synthmag\ was also used to determine \vs\ from the magnetically insensitive lines, such as Fe\i\ 5434.52~\AA. 

Table~\ref{params-table} summarises basic parameters of all program stars. In this table we provide an average \teff\ obtained with the two Str\"omgren photometric calibrations and the surface gravity obtained with the calibration by \citet{moon:1985}. It is known that photometric determination of the latter parameter often yields an overestimated surface gravity. Table~\ref{params-table} also provides an estimate of the projected rotational velocity and mean field modulus using the methods described above.

Time-averaged spectra for all program stars, including HD\,177765 \citep{alentiev:2012}, are presented in Fig.~\ref{allspectra}. Most objects are slow rotators, exhibiting rich spectra and, in particular, a prominent Nd\iii\ 6145~\AA\ line. Zeeman splitting of the Fe\ii\ 6149~\AA\ line is evident in several stars.

\section{Results}
\label{results}

Here we report results of the search of short-period oscillations in the target stars. The following sections present details on the analysis of the newly discovered roAp stars HD\,132205, HD\,148593 and HD\,151860 (Sect.~\ref{newroap}), the roAp stars known from previous photometric observations HD\,119027 and HD\,185256 (Sect.~\ref{confroap}), and confirmation of recent spectroscopic detections of pulsations in HD\,69013, HD\,96237, HD\,143487 (Sect.~\ref{knownroAp}). Null results for HD\,5823, HD\,178892 and HD\,185204 are presented in Sect.~\ref{noAp}. The outcome of the frequency analysis of individual ions is presented in Tables~\ref{tbl:newroAps}--\ref{tbl:specroAps}.

\subsection{New roAp stars}
\label{newroap}

\begin{table}%[!t]
\caption{Results of the frequency analysis for the newly discovered roAp stars. The columns give the stellar name, ion identification, the number of studied spectral lines and the amplitude $A$ and phase $\varphi$ derived by the least-squares cosine fit. The phase is given as a fraction of the averaged pulsation period indicated in the table. The false alarm probability for the variability detected in each ion is given in the last column.}
\label{tbl:newroAps}
\centering
\begin{tabular}{llccccc}
\hline
\hline
Star &Ion & N  & $A$ (\ms) & $\varphi$ & FAP \\
\hline
\multicolumn{6}{c}{$P=7.14\pm0.02$~min} \\
HD 132205 & Ce\ii & 10     & 95.7$\pm$8.0   & 0.92$\pm$0.03    & 3.2E-7     \\
 & Nd\iii     &  7     & 44.8$\pm$4.2    & 0.04$\pm$0.03    & 1.1E-6    \\
 & Pr\iii     &  6     & 51.7$\pm$5.7    & 0.93$\pm$0.04    & 7.1E-6    \\
 & Gd\ii      & 13     & 45.8$\pm$5.8    & 0.04$\pm$0.04    & 3.2E-5    \\
 & Sm\ii      &  5     & 97.4$\pm$13.9   & 0.04$\pm$0.03    & 1.4E-4    \\
 & Tb\iii     &  4     & 75.7$\pm$12.1   & 0.99$\pm$0.05    & 5.3E-4    \\
 & Dy\iii     &  4     & 76.4$\pm$12.2   & 0.99$\pm$0.05    & 6.2E-4    \\
 & H$\alpha$  & 1      & 90.5$\pm$15.6   & 0.83$\pm$0.05    & 1.4E-3    \\
 & Nd\ii      & 20     & 37.8$\pm$6.9    & 0.98$\pm$0.06    & 3.1E-3    \\
 & Er\ii      &  3     & 58.2$\pm$10.9   & 0.02$\pm$0.06    & 4.1E-3    \\
\hline
\multicolumn{6}{c}{$P=10.69\pm0.08$~min} \\
HD 148593 &  Nd\ii &  36 &     45.3$\pm$5.5 &    0.21$\pm$0.04  &     2.3E-5  \\
 &  Sm\ii  & 16   & 47.2$\pm$6.1    & 0.24$\pm$0.04    & 4.5E-5  \\
 &  Nd\iii & 12   & 23.8$\pm$4.1    & 0.31$\pm$0.05    & 1.5E-3  \\
 & Gd\ii   & 23   & 27.8$\pm$4.9    & 0.32$\pm$0.05    & 1.9E-3  \\
\hline
\multicolumn{6}{c}{$P=12.30\pm0.09$~min} \\
HD 151860 &    Tb\iii &   5 &       83.7$\pm$9.5   &  0.66$\pm$0.04     & 1.0E-5  \\
 & Eu\ii &   3 & 45.7$\pm$5.3    & 0.42$\pm$0.04   & 1.2E-5  \\
 & La\ii &  25 & 23.5$\pm$3.1    & 0.50$\pm$0.04   & 5.0E-5  \\
 & Dy\iii&   3 & 49.8$\pm$9.2    & 0.20$\pm$0.06   & 3.5E-3  \\
\hline
\end{tabular}
\end{table}

\subsubsection{HD 132205}
\label{hd132205}

The star HD\,132205 is classified as A2 EuSrCr in the catalogue by \citet{renson:2009}. \citet{martinez:1994} observed this star with time-resolved photometry in Johnson B filter during one hour on a single night, finding no pulsational variability above 0.5~mmag in the typical roAp frequency range. No other reports on this object are available in the literature.

\begin{figure}
\centering
\fifps{8cm}{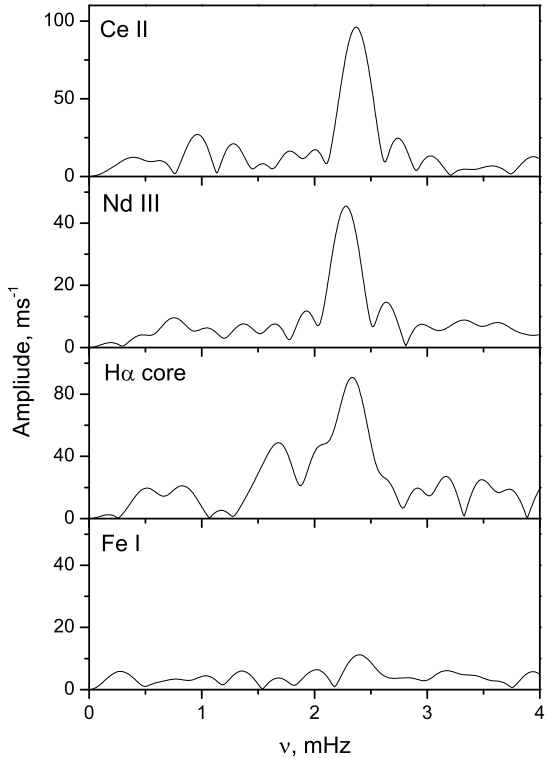}
\caption{Amplitude spectra of Ce\ii, Nd\iii, \ha\ core and Fe\i\ for HD\,132205.}
\label{132205}
\end{figure}

We have observed HD\,132205 with a time series of 50 UVES spectra. Analysis of these data immediately showed RV variability in many rare-earth element (REE) ions and in the H$\alpha$ core. This observation establishes HD\,132205 as a new roAp star. Fig.~\ref{132205} illustrates the amplitude spectra of Ce\ii, Nd\ii\ and \ha. This figure also shows the absence of significant variability in the Fe\i\ lines, confirming that oscillations detected in the REE lines are not due to an instrumental artifact. The amplitudes and phases of the mean RV curves of different ions are reported in Table~\ref{tbl:newroAps}. Using measurements derived from the Ce\ii, Gd\ii, Nd\iii\ and Pr\iii\ lines, we estimated an average pulsation period of $P=7.140\pm0.021$~min. The amplitude of oscillations does not exceed 100~\ms\ for all ions. Pulsation phases of different groups of variable spectral lines differ marginally. 

We determined \vs\,=\,9.5~\kms\ from a spectrum synthesis fit to the magnetically insensitive spectral lines. The projected rotational velocity of HD\,132205 is too large to directly detect magnetically split spectral lines in its spectrum. Nevertheless, spectrum synthesis analysis of the Fe\i\ 6336.8~\AA\ line gives evidence of a fairly strong magnetic field with \bs\,=\,5.2~kG.

\subsubsection{HD 148593}
\label{hd148593}

HD\,148593 is another cool Ap star for which practically no information is available in the literature. \citet{renson:2009} give A2 Sr spectral classification for this object. \citet{martinez:1994} failed to detect pulsational variability with 0.65~h photometric monitoring on a single night. \citet{wraight:2012} found no rotational photometric modulation based on observations with the STEREO satellites.

\begin{figure}
\centering
\fifps{8cm}{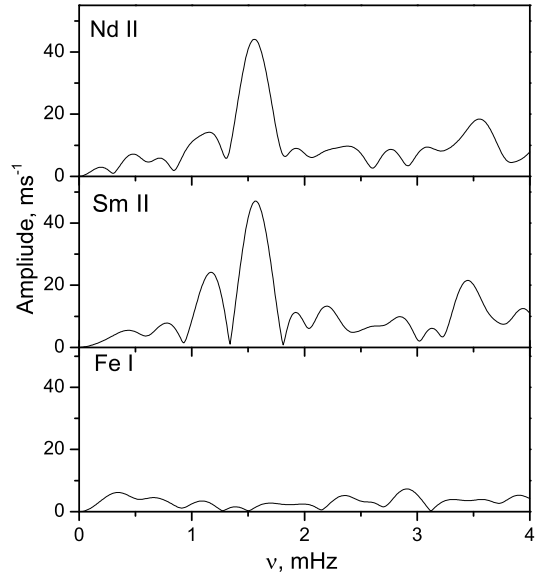}
\caption{Amplitude spectra of Nd\ii, Sm\ii\ and Fe\i\ for HD\,148593.}
\label{148593}
\end{figure}

Our UVES observations of this star consisted of 50 high-quality spectra. Radial velocity measurements show convincing evidence of relatively weak pulsations in several REE ions. Thus, HD\,148593 is definitely a roAp star. The most significant variability occurs in Nd\ii, Sm\ii, Nd\iii\ and Gd\ii. Using the two former ions we established a mean period of $P=10.690\pm0.081$ min. The amplitude spectra of Nd\ii\ and Sm\ii\ are compared with the measurements for constant Fe\i\ lines in Fig.~\ref{148593}. The complex shape of the REE amplitude spectra, especially for Sm\ii, suggests the presence of multiperiodic oscillations which are not resolved by our short time series data. The characteristics of the mean RV curves are reported in Table~\ref{tbl:newroAps}. All RV amplitudes are within 20--50~\ms\ range. We do not find a significant difference in the pulsation phases for variable spectral lines.

We used the magnetically insensitive Fe\i\ line to estimate \vs\,=\,5~\kms. There are no resolved Zeeman split lines in the spectrum of HD\,148593, but the magnetic field is likely to be present. We estimated \bs\,=\,3.0~kG from the spectrum synthesis fit using Fe lines with large Land\'e factors.

\subsubsection{HD 151860}
\label{hd151860}

Similar to the two previous objects, not much is known about the A2 SrEu star \citep{renson:2009} HD\,151860. It was observed in several photometric surveys of Ap stars \citep{hauck:1982,maitzen:1983,maitzen:2000} and was investigated for pulsations by \citet{martinez:1994}. These authors could not find photometric variability stronger than about 0.5~mmag from the 0.73~h observing run on a single night and hence classified this star as noAp.

\begin{figure}
\centering
\fifps{8cm}{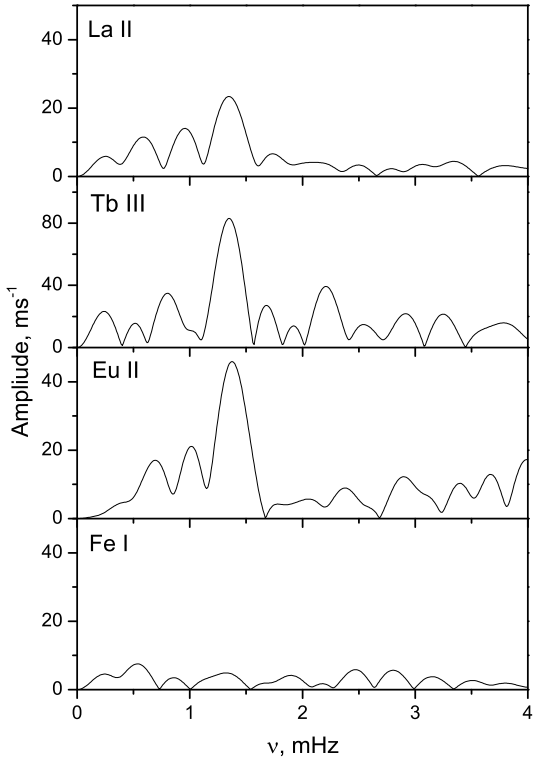}
\caption{Amplitude spectra of La\ii, Tb\iii, Eu\ii\ and Fe\i\ for HD\,151860.}
\label{151860}
\end{figure}

Our set of 50 UVES spectra of HD\,151860 shows definite pulsational variability in several REE ions, establishing this object as a new roAp star. The amplitude spectra of La\ii, Tb\iii\ and Eu\ii\ are presented in Fig.~\ref{151860}. In comparison, Fe\i\ lines show no significant oscillations. The highest amplitude peak occurs around 1.35~mHz for all REE ions. At the same time, several other peaks are present at lower frequencies. It is very likely that this object is a multiperiodic roAp star. An average pulsation period of $P=12.304\pm0.086$~min was established from the lines of  La\ii, Tb\iii\ and Eu\ii. All RV amplitudes are below 50~\ms\ except Tb\iii, for which oscillations reach 84~\ms. Pulsational phases differ significantly between REE ions. The mean RV curve of Dy\iii\ shows the most deviating behavior, lagging by $\sim$\,0.3 pulsation period behind other ions.

The projected rotational velocity \vs\,=\,4.5~\kms\ was determined from the Fe\i\ 5434.5~\AA\ line. Spectrum synthesis analysis of the magnetically sensitive spectral lines indicates a moderate \bs\ of around 2.5~kG.

\subsection{Known photometric roAp stars}
\label{confroap}

\begin{table}%[!t]
\caption{Results of the frequency analysis for the previously known photometric roAp stars. The columns are the same as in Table ~\ref{tbl:newroAps}.
Superscripts denote data sets obtained on different observing nights for HD\,119027.}
\label{tbl:photroAps}
\centering
\begin{tabular}{llccccc}
\hline
\hline
Star &Ion & N  & $A$ (\ms) & $\varphi$ & FAP \\
\hline
\multicolumn{6}{c}{$P=8.63\pm0.02$~min} \\
HD 119027$^1$ & Nd\ii  & 37      & 100.8$\pm$4.9   & 0.70$\pm$0.02     & 9.0E-9    \\
 & Ce\ii   & 30       & 97.6$\pm$4.9   & 0.76$\pm$0.02    & 1.3E-8    \\
 & Nd\iii  & 16      & 70.0$\pm$4.6    & 0.64$\pm$0.02     & 4.7E-8    \\
 & Pr\iii  & 9       & 66.2$\pm$4.9    & 0.46$\pm$0.02    & 1.2E-7    \\
 & Sm\ii   & 8       & 113.6$\pm$10.4  & 0.80$\pm$0.03    & 8.5E-7    \\
 & La\ii   & 28      & 33.6$\pm$5.7    & 0.00$\pm$0.05    & 1.4E-3    \\
 & Dy\iii  & 4       & 107.2$\pm$18.4  & 0.77$\pm$0.05    & 1.8E-3    \\
 & Gd\ii   & 11      & 74.4$\pm$13.0   & 0.72$\pm$0.05    & 2.0E-3    \\
\hline
\multicolumn{6}{c}{$P=8.76\pm0.01$~min} \\
HD 119027$^2$ & Nd\iii &16     &97.8$\pm$6.3   &0.21$\pm$0.02   &4.6E-8   \\
 & Nd\ii  &32     &118.7$\pm$8.8   &0.26$\pm$0.02   &1.0E-7   \\
 & Ce\ii  &31     &74.7$\pm$7.1    &0.39$\pm$0.03  &1.2E-6   \\
 & Pr\iii & 7     &77.0$\pm$7.5    &0.94$\pm$0.03  &1.5E-6   \\
 & Sm\ii  & 9     &136.1$\pm$14.9  &0.38$\pm$0.03  &6.1E-6   \\
 & Dy\iii & 4     &148.0$\pm$23.7  &0.33$\pm$0.05  &6.2E-4   \\
 & Gd\ii  &10     &66.1$\pm$13.2   &0.48$\pm$0.06  &1.0E-2   \\
 & La\ii  &26     &29.7$\pm$7.3    &0.33$\pm$0.08  &9.6E-2   \\
\hline
\multicolumn{6}{c}{$P=10.33\pm0.03$~min} \\
%HD 185256 & Ce\ii &  17 &     108.1$\pm$8.3 &    0.75$\pm$0.02 &     1.6E-7   \\
% & Dy\iii    & 5     & 174.7$\pm$14.2   & 0.72$\pm$0.02     & 2.6E-7  \\
% & Nd\iii    & 12    & 67.0$\pm$5.5     & 0.64$\pm$0.03     & 2.9E-7  \\
% & Sm\ii     & 6     & 125.8$\pm$12.0   & 0.92$\pm$0.03     & 1.0E-6  \\
% & Pr\iii    & 3     & 76.8$\pm$9.3     & 0.51$\pm$0.04     & 1.8E-5  \\
% & Nd\ii     & 17    & 75.1$\pm$10.8    & 0.68$\pm$0.04     & 2.6E-4  \\
% & Tb\iii    & 5     & 106.0$\pm$16.5   & 0.44$\pm$0.05     & 4.4E-4  \\
% & Gd2       & 33    & 35.0$\pm$6.8     & 0.77$\pm$0.06     & 1.1E-2  \\
% & H$\alpha$ & 1     & 129.8$\pm$28.9   & 0.85$\pm$0.07     & 2.7E-2  \\
%
HD 185256 & Ce\ii    & 7     & 133.3$\pm$11.7    & 0.71$\pm$0.03 & 5.1E-7 \\
 & Nd\iii   & 5     & 99.1$\pm$9.0      & 0.60$\pm$0.03 & 7.6E-7 \\
 & Sm\ii    & 4     & 126.7$\pm$12.1    & 0.81$\pm$0.03 & 1.5E-6 \\
 & Tb\iii   & 5     & 245.2$\pm$23.8    & 0.35$\pm$0.03 & 1.5E-6 \\
 & Dy\iii   & 3     & 247.0$\pm$29.0    & 0.70$\pm$0.03 & 1.1E-5 \\
 & Pr\iii   & 4     & 80.7$\pm$11.3     & 0.51$\pm$0.05 & 1.2E-4 \\
 & H$\alpha$& 1     & 129.8$\pm$28.9    & 0.85$\pm$0.07 & 2.7E-2 \\
 & Nd\ii    & 4     & 118.8$\pm$26.8    & 0.67$\pm$0.07 & 2.8E-2\\
\hline
\end{tabular}
\end{table}

\subsubsection{HD 119027}
\label{hd119027}

Photometric variability in A3 SrEu \citep{renson:2009} star HD\,119027 was discovered by \citet{martinez:1993a}. They reported five frequencies with a spacing of 26~$\mu$Hz clustered around 8.8~min pulsation period. A follow-up study by \citet{martinez:1998a} found evidence for two additional frequencies but could not unambiguously confirm the frequency spacing established in the previous study. Instead, the authors suggested that a more likely frequency spacing is 52~$\mu$Hz. Using mean light observations obtained during 45 individual nights, \citet{martinez:1998b} found no evidence of the rotational photometric variability of HD\,119027. They suggested that the rotational period may exceed 6~months or that the star is visible pole-on. \citet{mathys:1997b} discovered resolved magnetically split Fe\ii\ 6149~\AA\ line in the spectrum of HD\,119027 and measured a mean field modulus of 3.16~kG. The authors suggested a weak variability of \bs\ on the time scale of several weeks.

\begin{figure}
\centering
\fifps{8cm}{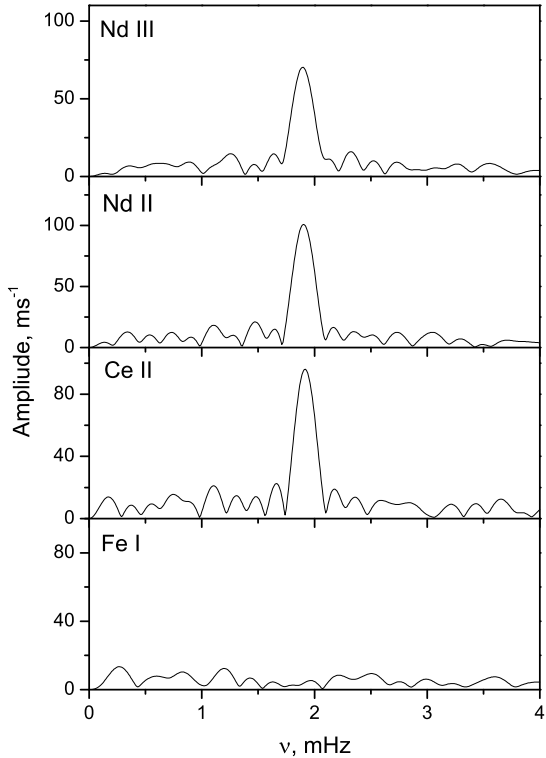}
\caption{Amplitude spectra of Nd\iii, Nd\ii, Ce\ii\ and Fe\i\ for HD\,119027 on the first observing night (data set 1).}
\label{119027_1}
\end{figure}

\begin{figure}
\centering
\fifps{8cm}{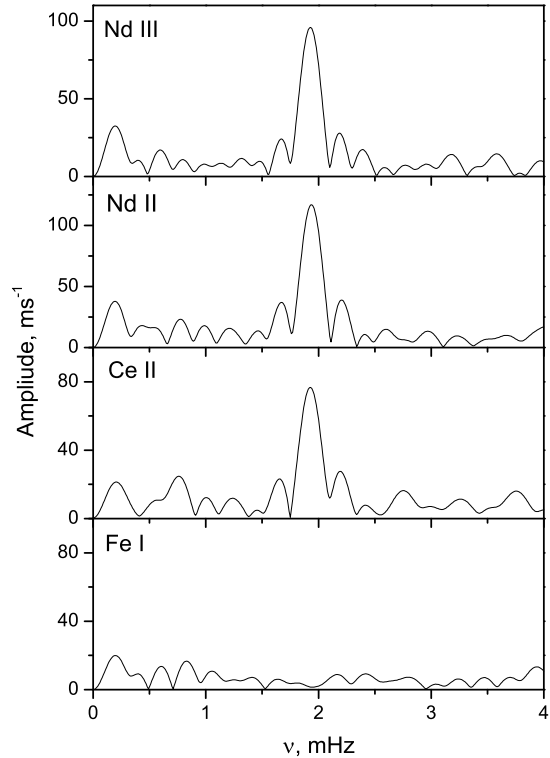}
\caption{Amplitude spectra of Nd\iii, Nd\ii, Ce\ii\ and Fe\i\ for HD\,119027 on the second observing night (data set 2).}
\label{119027_2}
\end{figure}

We obtained two UVES data sets for HD\,119027, each consisting of 50 spectra. Observations were separated by 11 nights. In both data sets a clear RV variation of the REE lines was found with a period similar to the previously known photometric one. This is the first report of the pulsational RV variability in this roAp star. The most secure identification of pulsations is found for Nd\iii, Nd\ii, Ce\ii, Pr\iii\ and Sm\ii. For all these ions pulsation amplitudes do not exceed 140~\ms. The amplitude spectra of Nd\iii, Nd\ii\ and Ce\ii\ are illustrated in Fig.~\ref{119027_1} and Fig.~\ref{119027_2} for the first and second data set respectively. The corresponding mean pulsation periods are $P_1= 8.627\pm0.016$ and $P_2=8.757\pm0.012$~min. The discrepancy between the period determinations for the two nights and a slightly different shape of the amplitude spectra are evident in Figs.~\ref{119027_1} and \ref{119027_2}. This is likely explained by the beating of close frequencies, which are not resolved in our short time series observations. HD\,119027 shows a large range of pulsation phases for different REE ions. Especially interesting is the behavior of Pr\iii, which exhibits a large phase lag relative to other ions. This may be an indication of pulsation node, similar to the one found in 33~Lib \citep{kurtz:2005} and 10~Aql \citep{sachkov:2008}.

For both observing nights we measured an identical field modulus of \bs\,=\,3.14~kG. There is no evidence of magnetic or spectroscopic variability between our two mean spectra. The projected rotational velocity of the star is estimated to be \vs\,=\,4~\kms.

\subsubsection{HD 185256}
\label{hd185256}

HD~185256 is the second known roAp star in our sample. Classified as F0 SrEu \citep{renson:2009}, it was discovered as pulsating by \citet{kurtz:1995}. Based on the photometric monitoring in the Johnson B filter, these authors found pulsations with a period of 10.2~min and an amplitude of 1.6~mmag. Apart from the brief communication by \citet{kurtz:1995}, no pulsational analysis of HD\,185256 has ever been published. The longitudinal magnetic field of \bz\,=\,$-706\pm180$~G was found in this star by \citet{hubrig:2004}. 

\begin{figure}
\centering
\fifps{8cm}{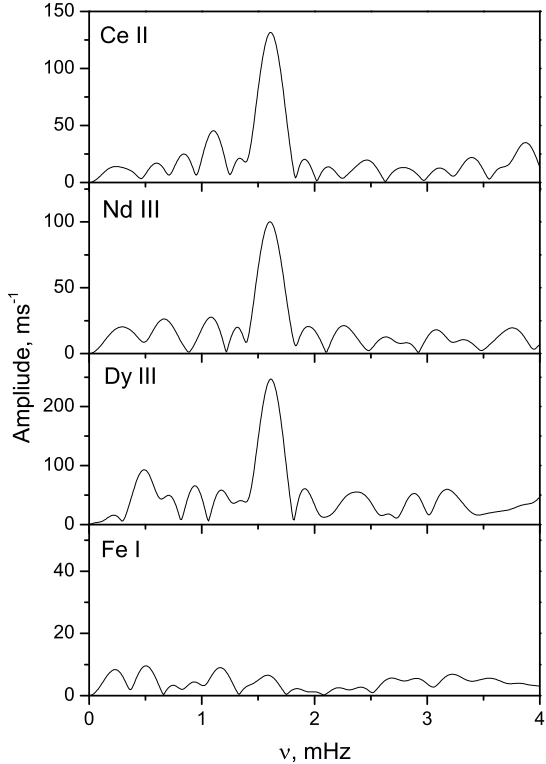}
\caption{Amplitude spectra of Ce\ii, Nd\iii, Dy\iii\ and Fe\i\ for HD\,185256.}
\label{185256}
\end{figure}

Analysis of our UVES spectra of HD\,185256 showed clear rapid RV variations in REE lines (Fig.~\ref{185256}). At the same time, the maximum pulsation amplitude in the Fe\i\ lines does not exceed 15~\ms. A marginal pulsation signal was detected in the \ha\ core and in Nd\ii\ lines. Radial velocity pulsations in HD\,185256 are typical of most roAp stars. The average period $P=10.333\pm0.029$~min was determined from the Ce\ii, Pr\iii, Nd\iii, Sm\ii, Tb\iii, and Dy\iii\ lines. The amplitudes of the oscillations vary between 80 and 250~\ms. The maximum amplitude is found for Dy\iii\ and Tb\iii\ lines. Pulsational phase shifts are also typical of the roAp stars, with the maximum RV first observed in singly-ionised REEs and H$\alpha$ core, followed by doubly-ionised REEs. The amplitude-phase relation for HD~185256 is similar to the one observed for another roAp star, 10~Aql \citep{sachkov:2008}. 

The projected rotational velocity of HD~185256, \vs\,=\,5.5~\kms, was estimated from the Fe\i~$\lambda$~5434~\AA\ line. The mean magnetic field modulus, \bb\,$\le$\,1.4~kG, was derived from the spectrum synthesis fit of the Fe\ii~$\lambda$~6149~\AA\ line. HD~185256 shows typical roAp-star abundance anomalies and the presence of a vertical chemical stratification in the atmosphere.      

\subsection{Known spectroscopic roAp stars}
\label{knownroAp}

\begin{table}%[!t]
\caption{Results of the frequency analysis for the confirmed new roAp stars. The columns are the same as in Table ~\ref{tbl:newroAps}.}
\label{tbl:specroAps}
\centering
\begin{tabular}{llccccc}
\hline
\hline
Star &Ion & N  & $A$ (\ms) & $\varphi$ & FAP \\
\hline
\multicolumn{6}{c}{$P=11.22\pm0.03$~min} \\
HD 69013 & Pr\iii & 9    &  73.0$\pm$4.6       & 0.88$\pm$0.02      & 8.0E-10 \\
 & Nd\iii & 13   &  97.3$\pm$6.6      & 0.94$\pm$0.02      & 1.6E-9 \\
 & Nd\ii  & 19   & 118.9$\pm$8.4      & 0.06$\pm$0.02      & 2.4E-9 \\
 & Er\ii  & 2    & 132.8$\pm$29.2     & 0.92$\pm$0.07      & 3.0E-2 \\
 & Eu\ii  & 2    & 60.6$\pm$14.6      & 0.22$\pm$0.07      & 5.5E-2 \\
\hline
\multicolumn{6}{c}{$P=13.89\pm0.04$~min} \\
HD 96237 & Nd\iii& 19&    142.0$\pm$5.1  &0.62$\pm$0.02    &2.9E-9  \\
 & Tb\iii&     7&    185.1$\pm$9.9   &0.52$\pm$0.02    &1.4E-8  \\
 & Nd\ii&      8&    148.9$\pm$10.1  &0.72$\pm$0.02    &6.1E-8  \\
 & Ce\ii&     25&    61.8$\pm$5.7    &0.71$\pm$0.03    &9.0E-7  \\
 & Tm\ii&      3&    110.8$\pm$12.2  &0.71$\pm$0.03    &6.2E-6  \\
 & Sm\ii&     10&    91.1$\pm$11.0   &0.79$\pm$0.04    &1.9E-5  \\
 & H$\alpha$&  1&    288.1$\pm$32.0  &0.71$\pm$0.03    &2.2E-5  \\
 & Pr\iii&     2&    90.9$\pm$13.6   &0.61$\pm$0.05    &2.8E-4  \\
 & Er\iii&     2&    103.6$\pm$15.9  &0.75$\pm$0.05    &3.5E-4  \\
 & Dy\ii&      2&    112.9$\pm$18.9  &0.63$\pm$0.05    &1.0E-3  \\
 & Er\ii&      3&    87.7$\pm$14.8   &0.75$\pm$0.05    &1.3E-3  \\
\hline
\multicolumn{6}{c}{$P=9.63\pm0.05$~min} \\
HD 143487 & Nd\ii & 30      & 29.8$\pm$3.8    & 0.02$\pm$0.04     & 9.1E-6 \\
 & Pr\iii    & 4     & 30.1$\pm$4.6    & 0.95$\pm$0.05   & 1.6E-4 \\
 & Nd\iii    & 12    & 25.4$\pm$4.3    & 0.08$\pm$0.05   & 7.4E-4 \\
 & H$\alpha$ & 1     & 109.2$\pm$19.0  & 0.90$\pm$0.06   & 9.3E-4 \\
 & Ce\ii     & 10    & 22.18$\pm$4.0   & 0.01$\pm$0.06   & 1.5E-3      \\
 & Sc\ii     & 3     & 26.4$\pm$4.9    & 0.94$\pm$0.06   & 2.6E-3 \\
\hline
\end{tabular}
\end{table}

\subsubsection{HD 69013}

The presence of oscillations in the A2 SrEu \citep{renson:2009} star HD\,69013 was tested by \citet{nelson:1993} and \citet{martinez:1994}. Both studies failed to detect pulsations stronger than 0.6--1~mmag based on a few 1--2~h-long photometric time series. The high-resolution spectrum of HD\,69013 obtained by \citet{freyhammer:2008} showed resolved Zeeman split lines corresponding to \bs\,=\,4.8~kG. The spectral appearance of HD\,69013 is typical of a roAp star. Using two short UVES spectroscopic time series data sets \citet{elkin:2011} found a RV variability with amplitudes of up to 200~\ms\ in individual REE spectral lines and a period of 11.4~min. Follow-up photometric observations, also published by \citet{elkin:2011}, suggested the presence of light variability with a similar period.

\begin{figure}
\centering
\fifps{8cm}{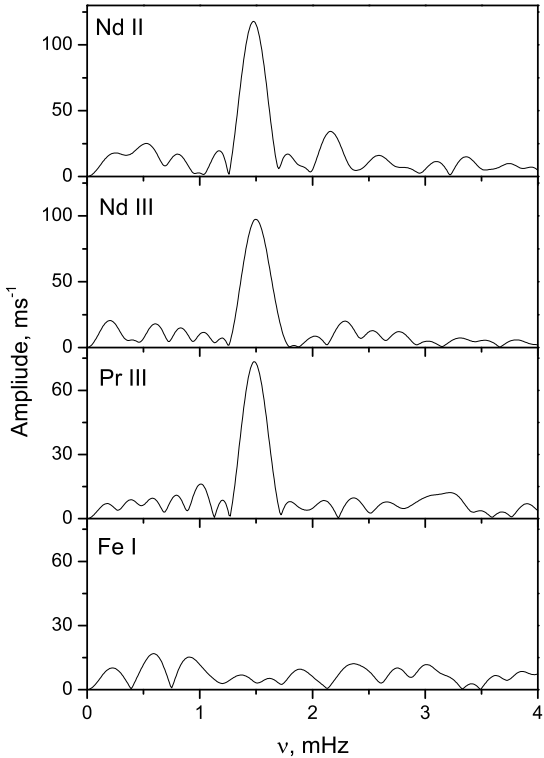}
\caption{Amplitude spectra of Nd\ii, Nd\iii, Pr\iii\ and Fe\i\ for HD\,69013.}
\label{69013}
\end{figure}

Our set of 62 spectra of HD\,69013 reveals unambiguous evidence of pulsations in several REE ions (Fig.~\ref{69013}). The RV amplitudes are typically 100~\ms\ or less. The most significant oscillations signals are found for the lines of Nd\ii, Nd\iii\ and Pr\iii. The signal-to-noise ratio of the corresponding peaks in the amplitude spectra is much higher than in the study by \citet{elkin:2011}. Using the mean RV curves of these REE ions, we determined the average pulsation period of $P=11.218\pm0.032$~min. The RV variations of these ions show a non-negligible phase shift. The RV maximum is first reached for Nd\ii, followed by Nd\iii\ and then Pr\iii. This order presumably reflects the relative formation heights of the lines of these ions.

Using a magnetically insensitive Fe\i\ line we measure \vs\,=\,3.5~\kms. On the other hand, the splitting of Fe\ii\ 6149~\AA\ indicates the mean field modulus \bs\,=\,4.8~kG. Both parameters are in excellent agreement with the results of \citet{freyhammer:2008}.

\subsubsection{HD 96237}
\label{hd96237}

A photometric search of pulsations in this A4 SrEuCr star \citep{renson:2009} was carried out by \citet{nelson:1993}. Based on 2-hour observations, they established an upper limit of 0.5 mmag on possible rapid oscillations. \citet{hubrig:2000b} treated this object as a non-pulsating star in their study of the H-R diagram position and kinematics of cool Ap stars. \citet{freyhammer:2008} detected Zeeman split lines indicating mean magnetic field \bs\,=\,2--3~kG and noted an unusually strong spectral variability in this star. They determined a rotation period of 20.91~d from the archival ground-based and space photometry. From a set of 34 time-resolved UVES spectra \citet{elkin:2011} found a marginal evidence of the RV variability with a period of $\approx$14~min and amplitudes up to 100~\ms\ in REE spectral lines.

\begin{figure}
\centering
\fifps{8cm}{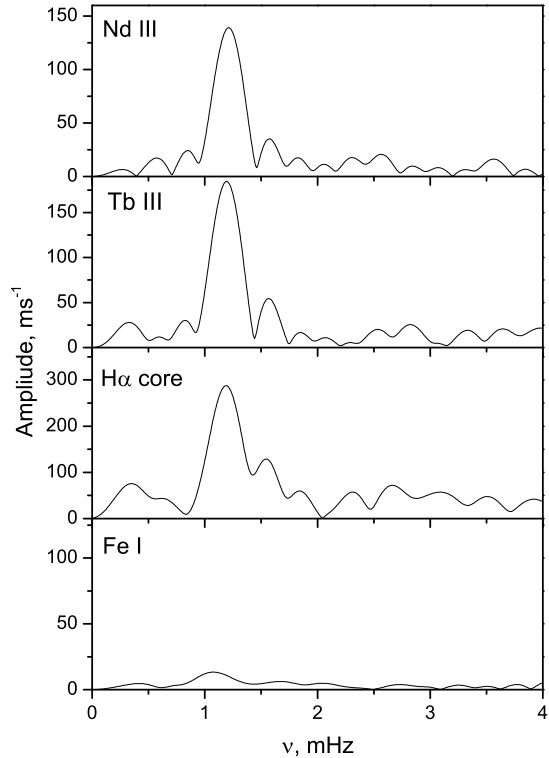}
\caption{Amplitude spectra of Nd\iii, Tb\iii, \ha\ core and Fe\i\ for HD\,96237.}
\label{96237}
\end{figure}

Our 50 UVES spectra of HD\,96237 show very clear pulsational signatures in different REE lines and in the core of H$\alpha$ (Fig.~\ref{96237}). The RV amplitudes exceed 100~\ms\ for many REE ions and reach 290~\ms\ in the core of H$\alpha$. The mean period, $P=13.866\pm0.036$ min, inferred from the measurements of Ce\ii, Tm\ii, Nd\ii, Nd\iii, Tb\iii\ and Dy\iii\ lines, is consistent with the results of \citet{elkin:2011}, yet our data yields detection of oscillations with 2--3 times higher signal-to-noise ratio than in the previous study. The least-squares analysis reveals a significant difference in phases of the RV curves of different groups of lines. Among the species with FAP\,$\le$\,$10^{-4}$ a larger phase (which corresponds to an earlier pulsation maximum) is found for H$\alpha$ and singly ionised REEs. On the other hand, Nd\iii\ and Tb\iii\ show a later maximum, suggesting formation of these lines in higher atmospheric layers.

Using spectrum synthesis modeling we found \vs\,=\,6~\kms\ and \bs\,=\,2.9~kG. Both parameters agree with the findings by \citet{freyhammer:2008}.

\subsubsection{HD 143487}
\label{hd143487}

This object is the coolest star in our sample. It is classified as A2 SrEuCr \citep{renson:2009}, but has a \teff\ around 7000~K according to our analysis. \citet{martinez:1994} could not find p-mode pulsations in this star stronger than $\approx$\,0.5--1.0~mmag based on the photometric observations obtained on three different nights. \citet{freyhammer:2008} detected resolved Zeeman split lines in the spectrum of HD\,143487 and identified this star as a promising roAp candidate based on its spectral similarity with known roAp stars. Using a small number of UVES spectra collected over 32~min, \citet{freyhammer:2008} suspected a RV variation at the frequency of 2~mHz. In a follow-up study \citet{elkin:2010a} examined three UVES data sets comprising 18--34 spectra and found variations with amplitudes 30--60~\ms\ and periods 8.8--10.0~min. This short time-series data did not allow \citet{elkin:2010a} to determine the main pulsation frequency with any more precision and to look for other modes.

Our UVES data set for HD\,143487 comprises 62 spectra collected over a time span of 1.4~h. Analysis of these spectra yields a highly significant detection of pulsations in the lines of Nd\ii, Nd\iii, Pr\iii, and in the core of H$\alpha$. Metal lines show amplitudes of 20--30~\ms, while pulsations reach 100~\ms\ in H$\alpha$. The average period, $P=9.631\pm0.053$ min, was determined from the lines of Nd\ii, Nd\iii\ and Pr\iii. Within the uncertainties of our analysis all groups of lines show the same pulsation phase. 

\begin{figure}
\centering
\fifps{8cm}{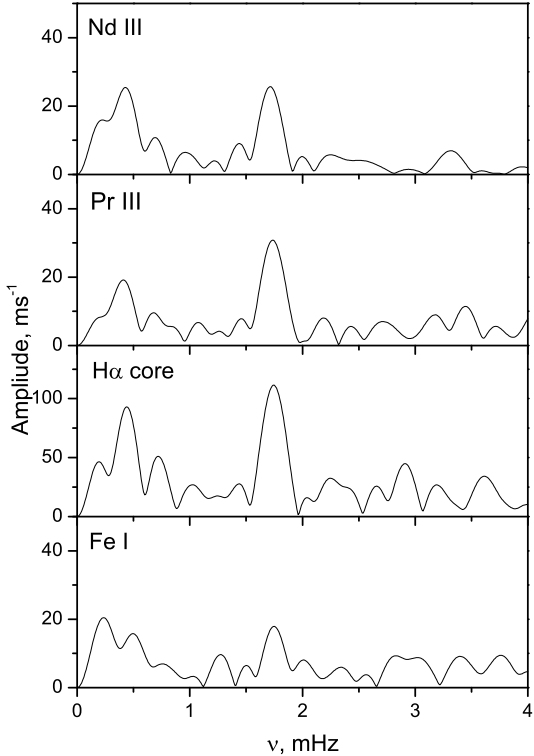}
\caption{Amplitude spectra of Nd\iii, Pr\iii, \ha\ core and Fe\i\ for HD\,143487.}
\label{143487}
\end{figure}

The representative amplitude spectra for HD\,143487 are shown in Fig.~\ref{143487}. The main pulsation peak is detected with a higher signal-to-noise ratio than in the study by \citet{elkin:2010a}. This figure also reveals a significant amplitude excess at low frequencies in all variable ions and even in the average RV curve of Fe\i. The corresponding variability with an amplitude of $\approx$\,20~\ms\ and a period of $\approx$\,42~min is not highly significant for any given ion but is reproduced for each group of lines. On the other hand, it is also present, with a somewhat lower amplitude, in the RV measurements of sharp telluric features around $\lambda=6300$ and 6900~\AA. Therefore, we tentatively attribute this low-frequency signal to an instrumental effect. HD\,143487 is the only star in our sample showing this artifact.

Using several REE lines with the triplet-like Zeeman splitting patterns, we determined \bs\,=\,$4.75\pm0.05$~kG, which noticeably exceeds \bs\,=\,4.2--4.3~kG found by \citet{freyhammer:2008}. We note that the Fe\ii\ 6149~\AA\ line on which Freyhammer et al. partly based their magnetic field measurements is strongly blended in the spectrum of HD\,143487 (see Fig.~\ref{allspectra}) and cannot provide a reliable estimate of \bs. On the other hand, our \vs\,=\,1.5~\kms\ is in reasonable agreement with their estimate of 2~\kms.

\subsection{Null results}
\label{noAp}

\subsubsection{HD 5823}
\label{hd5823}

HD\,5823 is classified as a F2 SrEuCr star \citep{renson:2009} and falls in the temperature range occupied by the roAp stars. Pulsations in this object have been repeatedly searched for in photometric ground-based surveys. But neither \citet{nelson:1993} nor \citet{martinez:1994} could detect pulsational variability. The latter authors observed HD\,5823 on three different nights, finding an upper photometric variability limit of 0.5--1.0~mmag.

As evident from Fig.~\ref{allspectra}, HD\,5823 shows relatively broad spectral lines compared to other stars in our sample. We measured \vs\,=\,13.5~\kms\ from the magneticaly insensitive Fe\i\ line.  Interestingly, this star also shows evidence of a fairly strong magnetic field. We find \bs\,=\,8.5~kG from the spectrum synthesis calculations. The Fe\ii\ 6149~\AA\ line shows a hint of partially resolved Zeeman splitting.

\begin{figure}
\centering
\fifps{8cm}{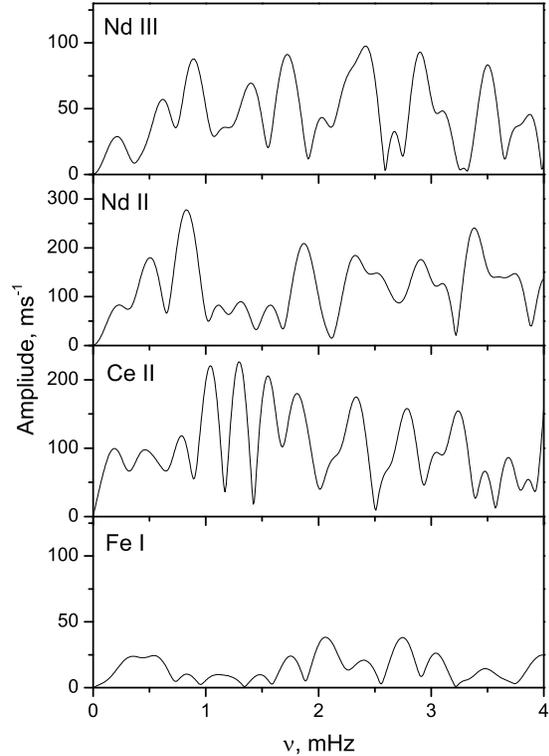}
\caption{Amplitude spectra of Nd\iii, Nd\ii, Ce\ii\ and Fe\i\ for HD\,5823.}
\label{5823}
\end{figure}

Due to a large line width, we could not reach the same precision in the RV analysis of HD\,5823 as in RV measurements for the other Ap stars. The amplitude spectra derived from the mean RV curves of Nd\iii\, Nd\ii\ and Ce\ii\ are shown in Fig.~\ref{5823}. The upper limit of RV oscillations is 100--200~\ms, which does not meaningfully constrain the roAp nature of this star. Weak pulsations, similar to those found for HD\,132205, HD\,148593, HD\,151860, could have been easily missed in our data for HD\,5823.

\subsubsection{HD 178892}
\label{hd178892}

This Ap SrCrEu star possesses one of the strongest magnetic fields among cool Ap stars. A strong longitudinal magnetic field, reaching 7.5~kG in maximum, was measured for HD\,178892 by \citet{kudryavtsev:2006}. \citet{ryabchikova:2006a} presented a detailed magnetic field, model atmosphere and abundance analysis of this star based on high-resolution spectra. They measured \bs\,=\,17.1--18.0~kG from the resolved magnetically split spectral lines and determined a rotational period of 8.2478~d from the ASAS photometry. A dipolar model fit suggested an inclination angle $i=37^{\circ}$ and a polar field strength $B_{\rm p}$\,$\approx$\,23~kG. Abundance analysis of this star performed by \citet{ryabchikova:2006a} revealed a typical roAp pattern. Based on these results, the authors suggested this star as an interesting target for the search of rapid oscillations. However, prior to our study, no time-resolved photometric or spectroscopic monitoring of this star was ever carried out.

\begin{figure}
\centering
\fifps{8cm}{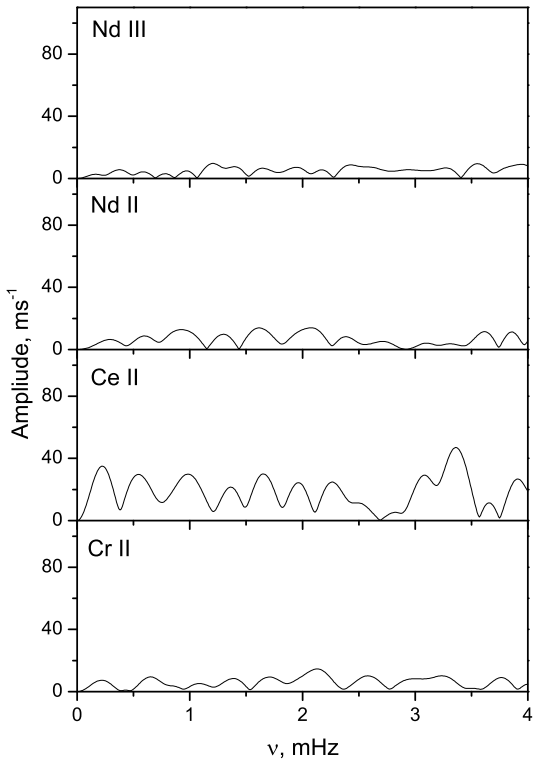}
\caption{Amplitude spectra of Nd\iii, Nd\ii, Ce\ii\ and Cr\ii\ for HD\,178892.}
\label{178892}
\end{figure}

We have acquired 67 time-resolved UVES spectra for HD\,178892. The analysis of rapid spectral line variability is complicated by a strong magnetic splitting of the majority of the lines. We have focused our RV measurements on the lines with small Land\'e factors and on the spectral features showing well-resolved Zeeman components. Despite a relatively high precision achieved for the mean RV curves of some REE ions, no oscillations were found. The amplitude spectra of Nd\ii, Nd\iii, Ce\ii\ and Cr\ii\ are illustrated in Fig.~\ref{178892}. The two former ions yield the upper limit of $\sim$\,10~\ms\ for the possible RV variability in the typical roAp frequency domain. This rules out the presence of pulsations with the amplitudes comparable to those found in other roAp stars in our survey.

The projected rotational velocity, \vs\,=\,10~\kms, estimated in our study is in a good agreement with \vs\,=\,$9\pm1$~\kms\ reported by \citet{ryabchikova:2006a}. We also determined a consistent mean field modulus of \bs\,=\,18.5~kG.

\subsubsection{HD 185204}
\label{hd185204}

HD\,185204 is another poorly studied cool Ap star with prominent chemical peculiarities and a strong magnetic field. It is classified as A2 SrEuCr \citep{renson:2009} and was investigated for pulsational variability by \citet{martinez:1994}. These authors obtained photometric observations on two nights, finding no oscillations exceeding $\sim$\,1~mmag. \citet{elkin:2012} detected resolved magnetically split lines in the spectrum of HD\,185204 and commented that this star is a promising target for searching for rapid oscillations. From the two spectra taken two nights apart \citet{elkin:2012} determined \bs\,=\,$5.7\pm0.2$~kG and \vs\,=\,4~\kms.

\begin{figure}
\centering
\fifps{8cm}{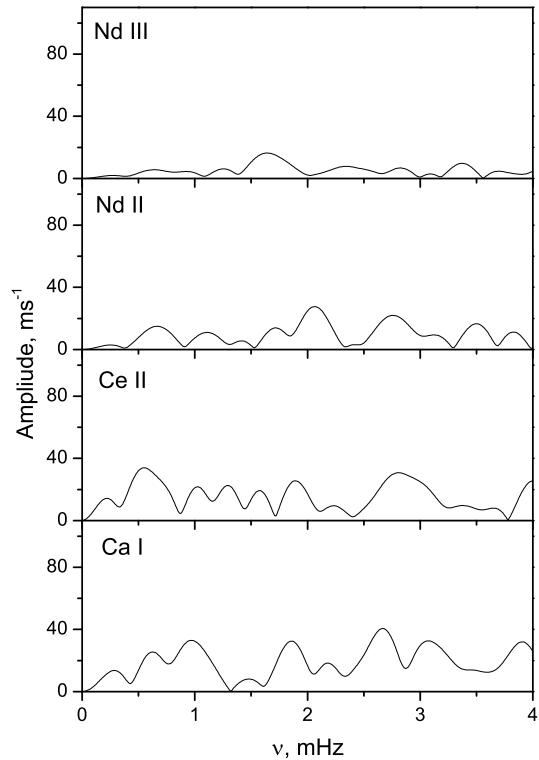}
\caption{Amplitude spectra of Nd\iii, Nd\ii, Ce\ii\ and Ca\i\ for HD\,185204.}
\label{185204}
\end{figure}

For this star we have two UVES data sets, one with 38 and another with 50 observations. We have focused RV analysis on the latter one. Due to the sharpness of its spectral lines, many REE absorption features can be measured in this star, resulting in accurate mean RV curves. In Fig.~\ref{185204} we illustrate the amplitude spectra for Nd\iii, Nd\ii, Ce\ii\ and Ca\i. There is no evidence of pulsational variability, with the upper RV amplitude limits of 20--40~\ms\ depending on the ion. Analysis of the first data set gave similar results, although the precision was a bit worse due to a smaller number of exposures. Thus, we spectroscopically confirm the noAp status of HD\,185204 to a good level of precision.

Our determination of the mean field modulus from the Zeeman-split lines yields \bs\,=\,5.4~kG. At the same time, the projected rotational velocity is found to be \vs\,=\,4.5~\kms. Given the precision of $\sim$\,0.1 for the magnetic field measurements, the difference with the \bs\ measurement by \citet{elkin:2012} is significant and possibly indicates a real variability corresponding to a long rotational period.

\section{Conclusions and discussion}
\label{conclusion}

In this paper we presented a spectroscopic search for high-overtone {\it p}-mode oscillations in cool magnetic Ap stars using the UVES spectrometer at the ESO 8-m VLT telescope. Our sample, comprising 12 objects, with 11 investigated here and another one studied by \citet{alentiev:2012}, was carefully chosen to satisfy several criteria. First, we have used the spectral classification information \citep[e.g.][]{renson:2009} to select Ap stars with anomalous Sr, Cr and Eu line strengths. Second, we have restricted the sample to have photometric \teff\ below $\sim$\,8000~K corresponding to the range where the roAp stars are found. Finally, we have examined the high-resolution spectra of every target, rejecting a number of stars with erroneous Ap classification or stars otherwise unsuitable for high-precision RV measurements (e.g. rapid rotators). Having established the stellar sample using this rigorous selection process, we were able to detect roAp pulsations in 9 out of 12 objects. This remarkable success rate makes our observing program the most effective search for roAp pulsators ever conducted.

Out of the 9 objects for which we have detected RV oscillations in high-resolution spectra, HD\,119027 and HD\,185256 were previously known photometric roAp stars but were lacking spectroscopic confirmations of pulsations. The remaining targets were classified, some of them repeatedly, as ``non-pulsating'' by the ground-based photometric observations \citep{nelson:1993,martinez:1994}. For three stars, HD\,69013, HD\,96237, HD\,143487, spectroscopic evidence of pulsations was independently presented by \citet{elkin:2010a,elkin:2011} using a lower quality spectroscopic material compared to our observations. We were able to confirm the presence of pulsational variability in these stars with much higher confidence levels. In addition, we discovered that four other cool Ap stars HD\,132205, HD\,148593, HD\,151860 and HD\,177765 are also roAp stars. The latter star, studied in detail by \citet{alentiev:2012}, is particularly interesting because it shows pulsations with a 24~min period -- longer than for any previously known roAp star.

Among the 9 objects for which we were able to detect spectroscopic pulsational variability, all stars showed oscillations in REE spectral lines. In addition, HD\,96237, HD\,132205, HD\,143487, HD\,177765 and HD\,185256 showed variability in the H$\alpha$ core. Typical pulsation amplitudes of new roAp stars are relatively low. The majority of stars exhibit RV oscillations below 100~\ms\ for most ions. HD\,132205 pulsates with the amplitudes below 50~\ms. Pulsation periods of newly discovered pulsators stars lie in the 7.1--13.9~min range, which is typical of roAp stars.

For three stars, HD\,5823, HD\,178892, HD\,185204, we were not able to detect spectroscopic oscillations. The relatively rapid rotation of HD\,5823 prevented us from obtaining precise RV measurements. Therefore, our upper limit on the possible pulsations in this star is not particularly stringent. For the other two stars we can rule out pulsations stronger than 10--20~\ms\ in the Nd\ii\ and Nd\iii\ lines. The extremely strong, 18.5~kG, magnetic field HD\,178892 makes it unusual among other roAp candidates. Only one roAp star, HD\,154708 \citep{kurtz:2006b}, with comparable magnetic field strength is known. Magnetic splitting of spectral lines in HD\,178892 and HD\,154708 complicates accurate RV measurements. Furthermore, the cool Ap stars with stronger fields seem to have systematically lower pulsation amplitudes compared to the weak-field roAp stars. We showed that the third non-pulsating star in our study, HD\,185204, possesses magnetic field of 5.4~kG, which is again significantly stronger than that found in a typical roAp star.

One of the goals of our survey was to answer the question of whether there is a substantial difference in the pulsational properties of the roAp stars and the cool Ap stars previously classified as non-pulsating based on photometric observations. We found weak oscillations in 7 objects previously classified as ``noAp''. It appears that the range of pulsational amplitudes of the roAp stars does not have a well-defined lower threshold but spans all the way from a few \kms\ in stars like HD\,83368 \citep{kochukhov:2006a} and HD\,99563 \citep{elkin:2005a} to our detection limit of $\sim$\,20~\ms. Bright roAp stars pulsating with amplitudes close to these limits are already known. For example, \citet{kochukhov:2009} detected 20--30~\ms\ multi-periodic pulsations in the Ap star HD\,75445. Similarly, \citet{kurtz:2007a} and \citet{kochukhov:2008} reported pulsations at the level of $\sim$\,20~\ms\ for $\beta$~CrB. Furthermore, observations with the Kepler satellite revealed roAp pulsations with amplitudes as low as few tens of $\mu$mag \citep{balona:2011}. All these results suggest that there is no real physical difference between the large-amplitude roAp stars and the so-called noAp stars. It is quite likely that all cool Ap stars are pulsationally unstable, and objects in which low-amplitude oscillations are currently being detected with spectroscopic monitoring represent the low-amplitude tail of the overall amplitude distribution.

\section*{Acknowledgments}
We thank James Silvester for improving the language of our paper.
OK is a Royal Swedish Academy of Sciences Research Fellow, supported by grants from Knut and Alice Wallenberg Foundation and Swedish Research Council.
DA and MC acknowledge financial support of FCT/MCTES, Portugal, through the project PTDC/CTE-AST/098754/2008. MC is partially funded by POPH/FSE (EC).
TR acknowledges support from Basic Research Program of the Russian Academy of Sciences ``Non-stationary phenomena in the Universe''.
WW was supported by the Austrian Science Fund (project P22691-N16).

%\bibliographystyle{mn2e}
%\bibliography{astro_papers}

\begin{thebibliography}{56}
\expandafter\ifx\csname natexlab\endcsname\relax\def\natexlab#1{#1}\fi

\bibitem[{{Alentiev} {et~al}\mbox{.}(2012){Alentiev}, {Kochukhov},
  {Ryabchikova}, {Cunha}, {Tsymbal}, \& {Weiss}}]{alentiev:2012}
{Alentiev} D., {Kochukhov} O., {Ryabchikova} T., {Cunha} M., {Tsymbal} V.,
  {Weiss} W., 2012, \mnras, 421, L82

\bibitem[{{Balmforth} {et~al}\mbox{.}(2001){Balmforth}, {Cunha}, {Dolez},
  {Gough}, \& {Vauclair}}]{balmforth:2001}
{Balmforth} N.~J., {Cunha} M.~S., {Dolez} N., {Gough} D.~O., {Vauclair} S.,
  2001, \mnras, 323, 362

\bibitem[{{Balona} {et~al}\mbox{.}(2011){Balona}, {Cunha}, {Kurtz},
  {Brand{\~a}o}, {Gruberbauer}, {Saio}, {{\"O}stensen}, {Elkin}, {Borucki},
  {Christensen-Dalsgaard}, {Kjeldsen}, {Koch}, \& {Bryson}}]{balona:2011}
{Balona} L.~A. {et~al.}, 2011, \mnras, 410, 517

\bibitem[{{Bigot} \& {Kurtz}(2011)}]{bigot:2011}
{Bigot} L., {Kurtz} D.~W., 2011, \aap, 536, A73

\bibitem[{{Cunha}(2002)}]{cunha:2002}
{Cunha} M.~S., 2002, \mnras, 333, 47

\bibitem[{{Elkin} {et~al}\mbox{.}(2005{\natexlab{a}}){Elkin}, {Kurtz}, \&
  {Mathys}}]{elkin:2005a}
{Elkin} V.~G., {Kurtz} D.~W., {Mathys} G., 2005{\natexlab{a}}, \mnras, 364, 864

\bibitem[{{Elkin} {et~al}\mbox{.}(2010){Elkin}, {Kurtz}, {Mathys}, \&
  {Freyhammer}}]{elkin:2010a}
{Elkin} V.~G., {Kurtz} D.~W., {Mathys} G., {Freyhammer} L.~M., 2010, \mnras,
  404, L104

\bibitem[{{Elkin} {et~al}\mbox{.}(2012){Elkin}, {Kurtz}, \&
  {Nitschelm}}]{elkin:2012}
{Elkin} V.~G., {Kurtz} D.~W., {Nitschelm} C., 2012, \mnras, 420, 2727

\bibitem[{{Elkin} {et~al}\mbox{.}(2011){Elkin}, {Kurtz}, {Worters}, {Mathys},
  {Smalley}, {van Wyk}, \& {Smith}}]{elkin:2011}
{Elkin} V.~G., {Kurtz} D.~W., {Worters} H.~L., {Mathys} G., {Smalley} B., {van
  Wyk} F., {Smith} A.~M.~S., 2011, \mnras, 411, 978

\bibitem[{{Elkin} {et~al}\mbox{.}(2005{\natexlab{b}}){Elkin}, {Riley}, {Cunha},
  {Kurtz}, \& {Mathys}}]{elkin:2005}
{Elkin} V.~G., {Riley} J.~D., {Cunha} M.~S., {Kurtz} D.~W., {Mathys} G.,
  2005{\natexlab{b}}, \mnras, 358, 665

\bibitem[{{Freyhammer} {et~al}\mbox{.}(2008){Freyhammer}, {Elkin}, {Kurtz},
  {Mathys}, \& {Martinez}}]{freyhammer:2008}
{Freyhammer} L.~M., {Elkin} V.~G., {Kurtz} D.~W., {Mathys} G., {Martinez} P.,
  2008, \mnras, 389, 441

\bibitem[{{Hatzes} \& {Mkrtichian}(2004)}]{hatzes:2004}
{Hatzes} A.~P., {Mkrtichian} D.~E., 2004, \mnras, 351, 663

\bibitem[{{Hauck} \& {North}(1982)}]{hauck:1982}
{Hauck} B., {North} P., 1982, \aap, 114, 23

\bibitem[{{Horne} \& {Baliunas}(1986)}]{horne:1986}
{Horne} J.~H., {Baliunas} S.~L., 1986, \apj, 302, 757

\bibitem[{{Hubrig} {et~al}\mbox{.}(2000){Hubrig}, {Kharchenko}, {Mathys}, \&
  {North}}]{hubrig:2000b}
{Hubrig} S., {Kharchenko} N., {Mathys} G., {North} P., 2000, \aap, 355, 1031

\bibitem[{{Hubrig} {et~al}\mbox{.}(2004){Hubrig}, {Szeifert}, {Sch{\"o}ller},
  {Mathys}, \& {Kurtz}}]{hubrig:2004}
{Hubrig} S., {Szeifert} T., {Sch{\"o}ller} M., {Mathys} G., {Kurtz} D.~W.,
  2004, \aap, 415, 685

\bibitem[{{Kaiser}(2006)}]{kaiser:2006}
{Kaiser} A., 2006, in Astronomical Society of the Pacific Conference Series,
  Vol. 349, Astrophysics of Variable Stars, {Aerts} C., {Sterken} C., eds., p.
  257

\bibitem[{{Khomenko} \& {Kochukhov}(2009)}]{khomenko:2009}
{Khomenko} E., {Kochukhov} O., 2009, \apj, 704, 1218

\bibitem[{{Kochukhov}(2004)}]{kochukhov:2004f}
{Kochukhov} O., 2004, \apjl, 615, L149

\bibitem[{{Kochukhov}(2006)}]{kochukhov:2006a}
{Kochukhov} O., 2006, \aap, 446, 1051

\bibitem[{{Kochukhov}(2007)}]{kochukhov:2007d}
{Kochukhov} O., 2007, in Physics of Magnetic Stars, {Romanyuk} I.~I.,
  {Kudryavtsev} D.~O., eds., pp. 109--118

\bibitem[{{Kochukhov}(2008)}]{kochukhov:2008c}
{Kochukhov} O., 2008, Communications in Asteroseismology, 157, 228

\bibitem[{{Kochukhov} {et~al}\mbox{.}(2009){Kochukhov}, {Bagnulo}, {Lo Curto},
  \& {Ryabchikova}}]{kochukhov:2009}
{Kochukhov} O., {Bagnulo} S., {Lo Curto} G., {Ryabchikova} T., 2009, \aap, 493,
  L45

\bibitem[{{Kochukhov} \& {Ryabchikova}(2001)}]{kochukhov:2001b}
{Kochukhov} O., {Ryabchikova} T., 2001, \aap, 374, 615

\bibitem[{{Kochukhov} {et~al}\mbox{.}(2008){Kochukhov}, {Ryabchikova},
  {Bagnulo}, \& {Lo Curto}}]{kochukhov:2008}
{Kochukhov} O., {Ryabchikova} T., {Bagnulo} S., {Lo Curto} G., 2008,
  Contributions of the Astronomical Observatory Skalnate Pleso, 38, 423

\bibitem[{{Kudryavtsev} {et~al}\mbox{.}(2006){Kudryavtsev}, {Romanyuk},
  {Elkin}, \& {Paunzen}}]{kudryavtsev:2006}
{Kudryavtsev} D.~O., {Romanyuk} I.~I., {Elkin} V.~G., {Paunzen} E., 2006,
  \mnras, 372, 1804

\bibitem[{{Kupka} {et~al}\mbox{.}(1999){Kupka}, {Piskunov}, {Ryabchikova},
  {Stempels}, \& {Weiss}}]{kupka:1999}
{Kupka} F., {Piskunov} N., {Ryabchikova} T.~A., {Stempels} H.~C., {Weiss}
  W.~W., 1999, \aaps, 138, 119

\bibitem[{{Kurtz} {et~al}\mbox{.}(2006){Kurtz}, {Elkin}, {Cunha}, {Mathys},
  {Hubrig}, {Wolff}, \& {Savanov}}]{kurtz:2006b}
{Kurtz} D.~W., {Elkin} V.~G., {Cunha} M.~S., {Mathys} G., {Hubrig} S., {Wolff}
  B., {Savanov} I., 2006, \mnras, 372, 286

\bibitem[{{Kurtz} {et~al}\mbox{.}(2005){Kurtz}, {Elkin}, \&
  {Mathys}}]{kurtz:2005}
{Kurtz} D.~W., {Elkin} V.~G., {Mathys} G., 2005, \mnras, 358, L6

\bibitem[{{Kurtz} {et~al}\mbox{.}(2007){Kurtz}, {Elkin}, \&
  {Mathys}}]{kurtz:2007a}
{Kurtz} D.~W., {Elkin} V.~G., {Mathys} G., 2007, \mnras, 380, 741

\bibitem[{{Kurtz} {et~al}\mbox{.}(2002){Kurtz}, {Kawaler}, {Riddle}, {Reed},
  {Cunha}, {Wood}, {Silvestri}, {Watson}, {Dolez}, {Moskalik}, {Zola},
  {Pallier}, {Guzik}, {Metcalfe}, {Mukadam}, {Nather}, {Winget}, {Sullivan},
  {Sullivan}, {Sekiguchi}, {Jiang}, {Shobbrook}, {Ashoka}, {Seetha}, {Joshi},
  {O'Donoghue}, {Handler}, {Mueller}, {Gonzalez Perez}, {Solheim},
  {Johannessen}, {Ulla}, {Kepler}, {Kanaan}, {da Costa}, {Fraga}, {Giovannini},
  \& {Matthews}}]{kurtz:2002}
{Kurtz} D.~W. {et~al.}, 2002, \mnras, 330, L57

\bibitem[{{Kurtz} \& {Martinez}(1995)}]{kurtz:1995}
{Kurtz} D.~W., {Martinez} P., 1995, Information Bulletin on Variable Stars,
  4209, 1

\bibitem[{{Kurtz} \& {Martinez}(2000)}]{kurtz:2000}
{Kurtz} D.~W., {Martinez} P., 2000, Baltic Astronomy, 9, 253

\bibitem[{{Maitzen} {et~al}\mbox{.}(2000){Maitzen}, {Paunzen}, {Vogt}, \&
  {Weiss}}]{maitzen:2000}
{Maitzen} H.~M., {Paunzen} E., {Vogt} N., {Weiss} W.~W., 2000, \aap, 355, 1003

\bibitem[{{Maitzen} \& {Vogt}(1983)}]{maitzen:1983}
{Maitzen} H.~M., {Vogt} N., 1983, \aap, 123, 48

\bibitem[{{Martinez}(1993)}]{martinez:1993}
{Martinez} P., 1993, PhD thesis, University of Cape Town, South Africa

\bibitem[{{Martinez} {et~al}\mbox{.}(1998{\natexlab{a}}){Martinez}, {Koen}, \&
  {Sullivan}}]{martinez:1998a}
{Martinez} P., {Koen} C., {Sullivan} D.~J., 1998{\natexlab{a}}, \mnras, 300,
  188

\bibitem[{{Martinez} \& {Kurtz}(1994)}]{martinez:1994}
{Martinez} P., {Kurtz} D.~W., 1994, \mnras, 271, 129

\bibitem[{{Martinez} {et~al}\mbox{.}(1993){Martinez}, {Kurtz}, \&
  {Meintjes}}]{martinez:1993a}
{Martinez} P., {Kurtz} D.~W., {Meintjes} P.~J., 1993, \mnras, 260, 9

\bibitem[{{Martinez} {et~al}\mbox{.}(1998{\natexlab{b}}){Martinez}, {Marang},
  {van Wyk}, \& {Roberts}}]{martinez:1998b}
{Martinez} P., {Marang} F., {van Wyk} F., {Roberts} G.~R., 1998{\natexlab{b}},
  The Observatory, 118, 153

\bibitem[{{Mathys} {et~al}\mbox{.}(1997){Mathys}, {Hubrig}, {Landstreet},
  {Lanz}, \& {Manfroid}}]{mathys:1997b}
{Mathys} G., {Hubrig} S., {Landstreet} J.~D., {Lanz} T., {Manfroid} J., 1997,
  \aaps, 123, 353

\bibitem[{{Mkrtichian} {et~al}\mbox{.}(2008){Mkrtichian}, {Hatzes}, {Saio}, \&
  {Shobbrook}}]{mkrtichian:2008}
{Mkrtichian} D.~E., {Hatzes} A.~P., {Saio} H., {Shobbrook} R.~R., 2008, \aap,
  490, 1109

\bibitem[{{Moon} \& {Dworetsky}(1985)}]{moon:1985}
{Moon} T.~T., {Dworetsky} M.~M., 1985, \mnras, 217, 305

\bibitem[{{Napiwotzki} {et~al}\mbox{.}(1993){Napiwotzki}, {Schoenberner}, \&
  {Wenske}}]{napiwotzki:1993}
{Napiwotzki} R., {Schoenberner} D., {Wenske} V., 1993, \aap, 268, 653

\bibitem[{{Nelson} \& {Kreidl}(1993)}]{nelson:1993}
{Nelson} M.~J., {Kreidl} T.~J., 1993, \aj, 105, 1903

\bibitem[{{Renson} \& {Manfroid}(2009)}]{renson:2009}
{Renson} P., {Manfroid} J., 2009, \aap, 498, 961

\bibitem[{{Ryabchikova} {et~al}\mbox{.}(2006){Ryabchikova}, {Kochukhov},
  {Kudryavtsev}, {Romanyuk}, {Semenko}, {Bagnulo}, {Lo Curto}, {North}, \&
  {Sachkov}}]{ryabchikova:2006a}
{Ryabchikova} T. {et~al.}, 2006, \aap, 445, L47

\bibitem[{{Ryabchikova} {et~al}\mbox{.}(2004){Ryabchikova}, {Nesvacil},
  {Weiss}, {Kochukhov}, \& {St{\"u}tz}}]{ryabchikova:2004a}
{Ryabchikova} T., {Nesvacil} N., {Weiss} W.~W., {Kochukhov} O., {St{\"u}tz} C.,
  2004, \aap, 423, 705

\bibitem[{{Ryabchikova} {et~al}\mbox{.}(2007{\natexlab{a}}){Ryabchikova},
  {Sachkov}, {Kochukhov}, \& {Lyashko}}]{ryabchikova:2007b}
{Ryabchikova} T., {Sachkov} M., {Kochukhov} O., {Lyashko} D.,
  2007{\natexlab{a}}, \aap, 473, 907

\bibitem[{{Ryabchikova} {et~al}\mbox{.}(2007{\natexlab{b}}){Ryabchikova},
  {Sachkov}, {Weiss}, {Kallinger}, {Kochukhov}, {Bagnulo}, {Ilyin},
  {Landstreet}, {Leone}, {Lo Curto}, {L{\"u}ftinger}, {Lyashko}, \&
  {Magazz{\`u}}}]{ryabchikova:2007a}
{Ryabchikova} T. {et~al.}, 2007{\natexlab{b}}, \aap, 462, 1103

\bibitem[{{Sachkov} {et~al}\mbox{.}(2008){Sachkov}, {Kochukhov}, {Ryabchikova},
  {Huber}, {Leone}, {Bagnulo}, \& {Weiss}}]{sachkov:2008}
{Sachkov} M., {Kochukhov} O., {Ryabchikova} T., {Huber} D., {Leone} F.,
  {Bagnulo} S., {Weiss} W.~W., 2008, \mnras, 389, 903

\bibitem[{{Saio} \& {Gautschy}(2004)}]{saio:2004}
{Saio} H., {Gautschy} A., 2004, \mnras, 350, 485

\bibitem[{{Saio} {et~al}\mbox{.}(2010){Saio}, {Ryabchikova}, \&
  {Sachkov}}]{saio:2010}
{Saio} H., {Ryabchikova} T., {Sachkov} M., 2010, \mnras, 403, 1729

\bibitem[{{Scargle}(1982)}]{scargle:1982}
{Scargle} J.~D., 1982, \apj, 263, 835

\bibitem[{{Th{\'e}ado} {et~al}\mbox{.}(2009){Th{\'e}ado}, {Dupret}, {Noels}, \&
  {Ferguson}}]{theado:2009}
{Th{\'e}ado} S., {Dupret} M.-A., {Noels} A., {Ferguson} J.~W., 2009, \aap, 493,
  159

\bibitem[{{Wraight} {et~al}\mbox{.}(2012){Wraight}, {Fossati}, {Netopil},
  {Paunzen}, {Rode-Paunzen}, {Bewsher}, {Norton}, \& {White}}]{wraight:2012}
{Wraight} K.~T., {Fossati} L., {Netopil} M., {Paunzen} E., {Rode-Paunzen} M.,
  {Bewsher} D., {Norton} A.~J., {White} G.~J., 2012, \mnras, 420, 757

\end{thebibliography}

\label{lastpage}

\end{document}